\documentclass[11pt, authoryear]{llncs}

\usepackage{natbib}
\usepackage{amsmath}
\usepackage{amssymb}
\usepackage{color}
\usepackage{enumerate}
\usepackage[margin=1in]{geometry}
\usepackage{bbm}
\usepackage{graphicx}

\usepackage{comment}

\newcommand{\E}{\mathop{\mathbb E}}

\DeclareMathOperator{\Bin}{Bin}

\newtheorem{observation}{Observation}

\begin{document}

\author{Amir Ban}
\institute{Hebrew University of Jerusalem}

\title{Unending Sequential Auctions}

\begin{titlepage}
\maketitle

\begin{abstract}
Sequential auctions for identical items with unit-demand, private-value buyers are common and often occur periodically without end, as new bidders replace departing ones. We model bidder uncertainty by introducing a probability that a bidder must exit the auction in each period. Treating the sequential auction as a Markov process, we demonstrate the existence of a unique steady state.

In the absence of uncertainty, the steady state resembles a posted-price mechanism: bidders with values above a threshold almost surely win items by repeatedly bidding the threshold price, while those below the threshold almost surely do not. The equilibrium price corresponds to the threshold value that balances supply (bidders with values above the threshold) and demand (auction winners).

When uncertainty is introduced, the threshold value persists but becomes less precise, growing ``fuzzier" as uncertainty increases. This uncertainty benefits low-value bidders—those below the threshold—by giving them a significant chance of winning. Surprisingly, high-value bidders also benefit from uncertainty, up to a certain value limit, as it lowers equilibrium bids and increases their expected utility. On the other hand, this bidder uncertainty often reduces the auctioneer's utility.
\end{abstract}

% Optionally include a table of contents
%\vspace{1cm}
%\setcounter{tocdepth}{2} % adjust to 1 if desired
%\tableofcontents

\end{titlepage}

\section{Introduction}

In sequential auctions several units are sold one after the other to the same group of buyers. The classical model (\cite{milgromweber2000, weber1981multiple}) describes such auctions where items are sold via first- or second-price {\em round} auctions to buyers with private values and unit demand. These auctions end when the last item is sold.

However, many real-world auctions, such as those for art, flowers, fish, wine, or satellite leases, operate over multiple sessions, with additional items available on subsequent dates. Buyers’ strategies in such cases are interdependent across sessions. Auctions in digital contexts, like keyword bidding on search engines or allocating cloud computing resources, often persist indefinitely. It is appropriate to call these sequential auctions {\em unending}, and analyze them as such.

Another example of an unending sequential auction is the competition among pending Bitcoin transactions to be recorded in a new Bitcoin block, and thus be executed, as well as become a permanent part of Bitcoin's ledger. The ``bid'' is the fee that the transaction promises the block ``miner'' for a place in the block. The miner's motivation to select the highest fees for inclusion makes this, since a block can hold many transactions, a multi-unit pay-your-bid auction akin to a first-price auction.

In this work, we study unending sequential auctions for identical items, sold to private-value buyers. A key question is how a buyer should behave in such a setting. Answering it is our main contribution. To answer it, we extend the classical model, while retaining its core assumptions. As background, here is a synopsis of classical model assumptions and results (\cite{krishna2009auction} Ch. 15):

A group of buyers compete for several identical items that are sold sequentially in first-price auctions, or alternatively, in second-price auctions. The buyers outnumber the items, and have private values, each independently drawn from a known distribution. Values are constant, per buyer, throughout the auction. The buyers have unit demand, so a round winner departs. The price paid is announced, but turns out to have no impact on future bidder strategies. 

In equilibrium, buyers bid a function of their value, a different function for each round, each conjectured to be increasing and differentiable. Consequently, in every round, the remaining bidder with the highest value wins. The equilibrium solution turns out to be based on the {\em order statistics} of a random sample of $(n-1)$ bidder values ($n$ being the number of bidders). Bids {\em increase} from round to round, as the number of bidders decreases, but the price of an item is a martingale, with constant expectation. Thus far the description of the classical model.

Returning to our unending auctions, we restrict ourselves at the outset to a first-price auction in each round, only briefly discussing second-price rounds (see Section \ref{second-price}). Any generalization of finite sequential auctions to infinite must include the arrival of new bidders, to at least offset the departure of round winners. We assume a Poisson arrival process of new bidders, with mean $\lambda$ per round. As we later show, only auctions with $\lambda > 1$ are non-trivial.

We introduce what we term {\em bidder uncertainty} to more realistically model bidder situations in unending auctions. We define bidder uncertainty as exogenous or endogenous factors that affect a bidder's valuation for an item, or her very presence as a buyer. No such element existed in the original classical model. However, further works described variations of the classical model with uncertainty. \cite{kittsteiner2004declining} added uniform (per bidder) discounting of value. %We quote from their work a partial list
% of real-world scenarios where their model of discounting may fit:
%``{\em This is the case if otherwise identical objects become available at different points in time and bidders prefer to receive an object early. Examples are auctions of goods on a rental basis where one object is available to one buyer at a time (like vacation accommodation, cars, tractors or DVD’s) or fish auctions, where the fish is auctioned when it arrives at the port. \ldots''} 

Another form of uncertainty for bidders is being limited in the number of rounds they participate in. Bidders may have a self-imposed deadline, or an exogenously imposed lifetime (a `buy-by' date, so to speak). In Bitcoin, for example, transactions that remain in the {\em mempool}, the repository of pending transactions, for 72 hours or more, are removed.

Our modeling choice for bidder uncertainty is the following: There is a uniform probability $\delta \geq 0$ that, independently, each round, each bidder is removed, or ``sent home''. The outcome is that bidders face a uniform and limited lifetime, geometrically distributed, with the past irrelevant for the future. We chose this uncertainty model since it is stateless and memoryless.

In unending auctions with {\em no uncertainty} bidders have unlimited lifetimes with unvarying valuations. We divide the bulk of our analysis between auctions with no uncertainty and auctions with uncertainty. The former has a main result, that the unending auction emulates a posted-price mechanism. The latter, with uncertainty, shows that such auctions approximate the no-uncertainty auction behavior, but in unexpected and economically-significant ways.

The typical solution concept for sequential auctions is a symmetric sequential equilibrium, and so is ours. But a finite auction solution consists in a {\em different} bidding function for each of its rounds, where the remaining bidders are aware of the number of rounds left. Our inquiry seeks a different sort of solution, for several reasons. The number of rounds left is always infinite, so in this respect at least, all rounds are the same. Furthermore, the starting point and history, at least the distant history, of an unending auction should be irrelevant (the Markov property). We expect the unending auction to reach (or limit at) a ``steady state'', by which we mean that all auction rounds are statistically equivalent. This is the target of our investigation.

More specifically, we look for a bidding strategy that is {\em unvarying} between rounds. The bidding strategy we seek takes the form of a function of the bidder value, which is bid repeatedly and consistently by each bidder, and, collectively, forms a symmetric sequential equilibrium.

This modeling decision is justified by the fact that we can, in fact, find such a ``steady-state'' equilibrium. This is achieved by modeling the unending sequential auction as a Markov process. Conveniently, Markov processes have a {\em stationary distribution} over their state space, under requirements that our model fulfills. Importantly, such a stationary distribution is {\em unique}. This unique stationary distribution serves as our sought-for ``steady state''. Deriving from it an unvarying bidding function, and other attributes of the equilibrium, is a technical task.

Caveats apply. The unending sequential auction is not necessarily in its stationary distribution. At best, that is its limiting distribution, approximating it as closely as required with time. Also, processes that do not have the Markov property, where history {\em is} relevant, do not necessarily have a stationary distribution. Recently, \cite{nisan2023serial} demonstrated unending oscillation of prices in a cryptocurrency context. Furthermore, price announcements of previous rounds can be relevant history. In the classical model, they are shown to have no effect on bidder behavior. This only partially holds in our context. Finally, the stationary distribution persists so long as the model parameters are constant. If, for example, there are long-term variations in the rate of new bidder arrival $\lambda$, the auction will drift to a different steady state.
 
We now describe our results.

The no-uncertainty unending auction is trivial if the bidder arrival rate has mean $\lambda \leq 1$ (or $\lambda \leq \mu$ for multi-unit auctions with $\mu$ winners) because, with patience, every bidder is assured a unit for free, by bidding `zero' repeatedly. On the other hand, if an unending auction has $\lambda > 1$, it must have a bidder pool size growing beyond any bound almost surely. This means it has no stationary distribution. %\footnote{over, e.g., the state space of the number of bidders in the bidder pool.}
However, we show that by excluding all bidders except those with sufficiently high values, the remaining bidders do have a stationary distribution, and their number is bounded. The upshot is that such an unending auction, in its steady state, emulates a {\em posted price} mechanism:
\begin{itemize}
\item The posted price is the $\frac{\lambda - 1}{\lambda}$ percentile of the bidder value distribution.
\item Bidders with value below the posted price, called {\em low-value} bidders, will almost surely never win an item. Low-value bidders bid their values in equilibrium.
\item Bidders with value above the posted price, called {\em high-value} bidders, will almost surely win an item. High-value bidders bid the posted price in equilibrium.
\end{itemize}

\begin{figure}[tbp]
    \centering
   % \begin{minipage}{.7\textwidth}
      \centering
		\includegraphics[height=0.3\textheight]{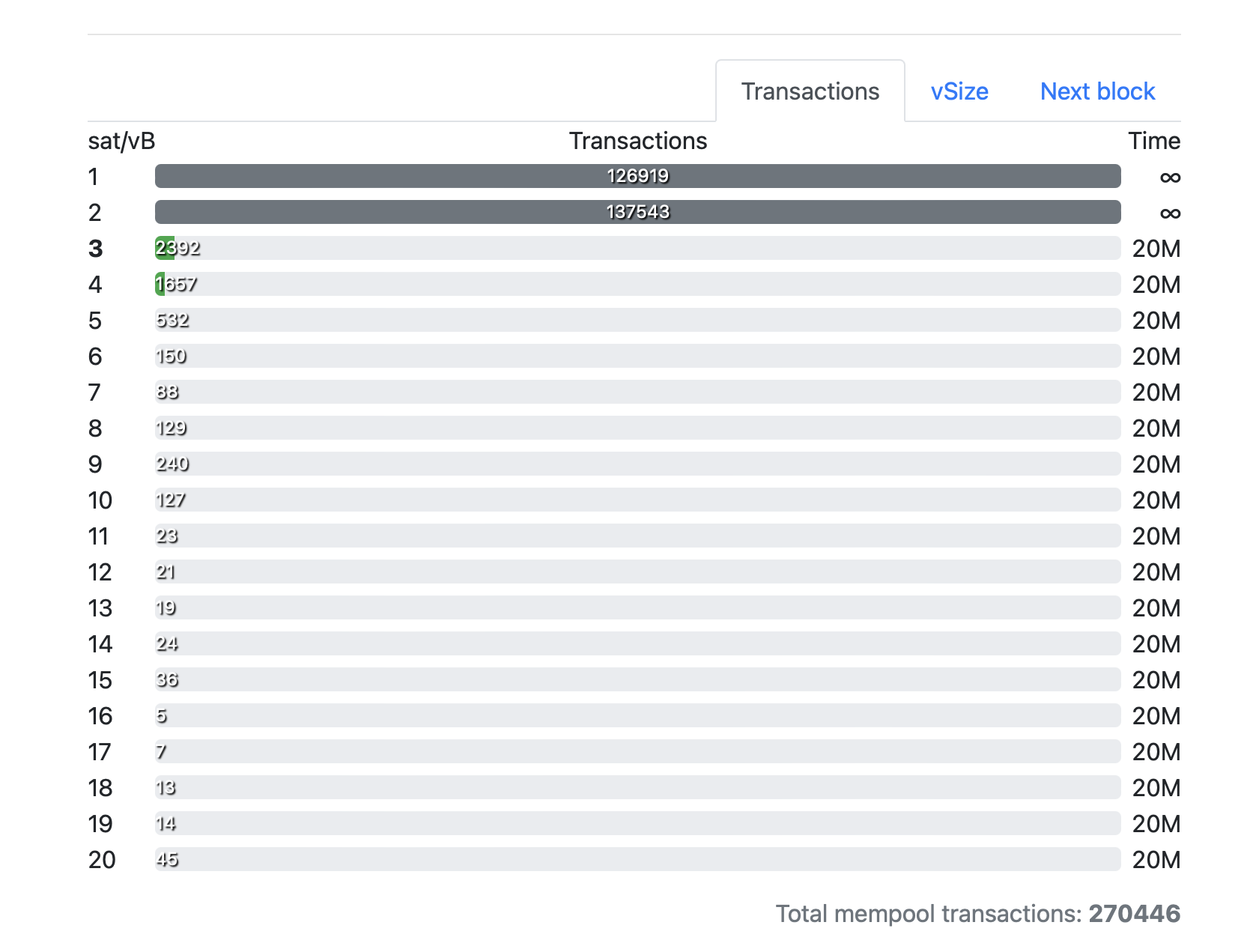}
		\caption{A snapshot of transaction fees in the mempool with expected waiting times (from \texttt{bitcoinfees.net})}
		\label{mempool}
    %\end{minipage}
%    \begin{minipage}{.49\textwidth}
%      \centering
%		\includegraphics[height=0.3\textheight]{bid5-u}
%		\label{bid5-u}
%    \end{minipage}
    \end{figure}

The price paid for an item in the no-uncertainty unending auction is, of course, the posted price. Price announcements are irrelevant to bidder strategies. An actual snapshot of the Bitcoin mempool, shown in Figure \ref{mempool}, is a good fit for our posted-price result. 

%We show that high-value bidders win an item in a geometrically distributed number of rounds, whose mean decreases with value, and diverges to infinity when their value approaches the posted price.

Turning to unending auctions with uncertainty, the analysis shows that, no matter how small $\delta$ is, the expected pool size is finite, equaling $O(\frac{\lambda}{\delta})$.
Our main result here is that uncertainty {\em favors} many bidders: It favors {\em all} low-value bidders, and all high-value bidders with sufficiently low values (i.e., only very-high-value bidders are worse off). Uncertainty is favorable to those bidders in comparison to the no-uncertainty case ($\delta=0$), and also, for sufficiently small $\delta$, its favorability is monotonic with $\delta$. The favorability, whenever it exists, is shown for two key bidder properties:
\begin{enumerate}
\item Bidders bid and pay less: Since they bid lower, their cost upon winning is lower. 
\item Expected utility is higher: Even with the risk of removal, their average payoff improves.
\end{enumerate}

Furthermore, we generalize these favorability-of-uncertainty results to {\em any} model of uncertainty, comprising any value-discounting schedule or restricted lifespan, whether deterministic or probabilistic.

While bidder uncertainty generally benefits buyers, its impact on the auctioneer is less clear. Although the game is not strictly zero-sum, it is likely that bidder gains translate into reduced auctioneer utility.

\section{Literature Review}
According to \cite{krishna2009auction}, equilibrium solutions of the classical model, with symmetric independent private values, which were first published in \cite{milgromweber2000}, were already derived by the same authors in 1982. \cite{weber1981multiple} treated a variation, with interdependent values.

\cite{kittsteiner2004declining} analyzed a model with uniform discounting of values per round (thus retaining the {\em highest-value-wins} property) in a finite sequential auction. By our definition, this exemplifies a model of bidder uncertainty. Its analysis in our unending context is probably intractable.

Dynamic auctions are auctions that change over time (\cite{lavi2004competitive} and many others). \cite{said2011sequential}, \cite{lavi2014efficiency}, have investigated unending random arrivals of new buyers and new items. However, these works are not focused on finding a steady state. \cite{che2025optimal} discuss many issues common with ours, but their results do not overlap with ours.

The Bitcoin protocol is described in detail in, e.g., \cite{nakamoto2008peer, chaudhary2015modeling, easley2019mining}. \cite{ferreira2021dynamic} proposed a posted-price mechanism  for Ethereum, but as an imposed rule, rather than our result of an organic, emergent phenomenon.

The rest of this paper is organized as follows: In Section \ref{model} we describe our model. In Section \ref{markov} we analyze our model with zero uncertainty and show the emergent posted-price mechanism. In Section \ref{uncertainty} we analyze the model with uncertainty. Concluding remarks are offered in Section \ref{discussion}.  Proofs and figures are found in the Appendix.

\section{Model}
\label{model}

A sequential auction is held periodically, without end, for identical items. Each round, one item is sold by a first-price auction.

There is a pool of bidders, which may hold any number of bidders at any time. Each bidder has a private value for an item, independently drawn from a known continuous distribution $X$, with c.d.f. $F(x)$ and density $f(x)$. $X$'s support infimum is $\underline{X} \geq 0$. A bidder's value is constant throughout being in the bidder pool.

The bidders have unit demand, so a bidder who wins an item leaves the pool. On the other hand, there is a Poisson-distributed arrival of new bidders each round, whose number is independently sampled from a Poisson distribution with expectation $\lambda$. In other words, with probability $e^{-\lambda} \frac{\lambda^m}{m!}$ the number of arrivals each round equals $m$, for all integer $m \geq 0$.
%\footnote{The Poisson distribution may be replaced by any other distribution with support over all the integers. All results, except perhaps Proposition \ref{mean}, would remain unchanged.} 
New arrivals have value independently drawn from the same distribution $X$ as older bidders.

We consider scenarios where the auctioneer does not announce the price for each unit, as well as when the auctioneer does announce prices. %In the latter case, buyers only know prices announced in rounds in which they competed.

We model bidder uncertainty by a spontaneous removal process, in which, every round, after the winner departs, and before new bidders arrive, each remaining bidder in the pool is permanently removed from the pool with probability $\delta \in [0,1]$. Such a removal event is `memoryless', and is independent of all previous or concurrent events. $\delta$ is a global model parameter, applied uniformly to all bidders. The higher $\delta$, the higher the uncertainty. $\delta = 0$ means there is no uncertainty, bidders are never removed, and, unless they win a round, remain in the pool indefinitely.

We seek a symmetric sequential equilibrium that is in a `steady state', meaning that every bidder bids {\em consistently} and {\em repeatedly} a function of her value, for as long as she participates in the auction, and cannot improve her utility by changing her bid in any or all rounds.
\footnote{An exception must be made for the situation where (i) A bidder is alone, or almost-alone in the bidder pool, and (ii) the bidder is aware of that. In such a case, a temporary deviation to a `zero' bid may be justified, in the hope of meeting no competitors. In most auctions, this is a very low-probability event.}

A standard assumption made in the analysis of sequential auctions is that bidding strategies in every round are (weakly) increasing in the bidder's value and are piecewise-differentiable. Consequently, the next item in the auction will be sold to the bidder with the highest value in the pool. Equal bids are tie-broken by value.

\section{The Zero-Uncertainty Unending Auction}
\label{markov}

We analyze the zero-uncertainty ($\delta=0$) model using the methodology of Markov chains to describe the dynamics of the bidder pool (See, for example, \cite{lalley2016markov} for a background on Markov chains and their terminology).

Let $N_t$ be the number of bidders in the pool at round $t$. Let $\mathcal{N}$ be the countable state space $N_t = 0, 1, 2, \ldots$. Let $\Lambda_t$ be the Poisson-distributed number of new bidders after round $t$. Then $(N_t)_{t \geq 0}$ is a Markov chain over state space $\mathcal{N}$ defined by the relation 
\begin{align}
\label{relation}
N_{t+1} = (N_t - 1)_+ + \Lambda_t
\end{align}

Define $S_t = \sum_{\tau=1}^t (\Lambda_t - 1)$ and $A_t = \sum_{\tau=1}^t \mathbbm{1}_{\{N_\tau = 0\}}$. Then 
\begin{align}
\label{N_t}
N_t = N_0 + S_t + A_t
\end{align}

By the Law of Large Numbers $\lim \inf S_t / t = \lambda - 1$, and so

\begin{proposition}[$\lambda > 1$]
\label{lambda>1}

If $\delta = 0$ and $\lambda > 1$, the pool size grows without limit: $\lim \sup N_t = +\infty$. I.e., there is a positive probability that, starting from $N_t = k$, $N_T \neq 0$ for every $T > t$, and this probability limits at $1$ as $k \to \infty$.
\end{proposition}
%\noindent i.e., there is a non-zero probability that the pool will never be empty, and this is almost surely true as the starting size grows.

On the other hand if $\lambda < 1$

\begin{proposition}[$\lambda < 1$]
\label{lambda<1}

If $\delta = 0$ and $\lambda < 1$ there is a unique stationary distribution over state space $\mathcal{N}$, i.e., a distribution that is unchanged by the chain relation \eqref{relation}. Furthermore, in the stationary distribution, the probability of an empty pool is $1 - \lambda$.
\end{proposition}

The case $\lambda = 1$ is borderline.
\begin{proposition}[$\lambda=1$]
\label{lambda=1}

If $\delta=0$ and $\lambda=1$, the chain is null-recurrent, i.e., from any starting point, $N_t = 0$ will occur almost surely, though the waiting time for such an event is unbounded.
\end{proposition}

Propositions \ref{lambda>1} thru \ref{lambda=1} show our investigation of unending auctions is non-trivial only for $\lambda > 1$. For $\lambda \leq 1$, a bidder can win a round, and the sought unit, for a price of $0$ (not even $\underline{X}$!) by bidding $0$, and waiting for an empty pool, herself excluded, when she has no competitors. By Propositions \ref{lambda<1} and \ref{lambda=1}, this will happen almost surely.
So the discussion of our subject, if it is not to be trivial, is left with $\lambda > 1$, and an ever-growing bidder pool, that will exceed any bound almost surely.

Nevertheless, Proposition \ref{lambda<1} remains highly relevant, due to the following observation.

\begin{observation}
\label{ob}

If the expected number of new arrivals per round is $\lambda$, each a random draw with value c.d.f. $F(x)$, then the expected number of new arrivals with value $>x$ is also Poisson distributed, with mean $\lambda [1 - F(x)]$. 
\end{observation} 

From this, one of our main results follows immediately: 
Define $X_{(\lambda)}$ to be the value for which $F(X_{(\lambda)}) = \frac{\lambda - 1}{\lambda}$, i.e., the value at the $\frac{\lambda - 1}{\lambda}$ percentile of $X$'s distribution. Given $\lambda$, we call bidders with value less than $F(X_{(\lambda)})$ {\em low-value} bidders, and those with value more than $F(X_{(\lambda)})$ {\em high-value} bidders. Then
\begin{theorem} [Winners' Threshold]
\label{threshold}

With $\lambda > 1$ and $\delta = 0$, high-value bidders will almost surely win a round, while low-value bidders will almost surely never win a round.
\end{theorem}

\begin{remark}
\label{rem}
Following Observation \ref{ob} we are following a continuum of Markov chains at once, each with a different parameter, the percentile of a bidder's value $F(x)$, marked $g := F(x)$. Accordingly, for every $g \in [0,1]$ there is a state space $\mathcal{N}_g$ of $N_t(g) = 0,1, 2, \ldots$, the number of bidders with value $> \lambda[1-g]$, and $(N_t(g))_{t\geq0}$ is a Markov chain over $\mathcal{N}_g$. %The state space of all bidders is $\mathcal{N}_0$.
\end{remark}

We now state the equilibrium bidding strategy (without uncertainty).
\begin{theorem}[Bidding without Uncertainty]
\label{posted-price}

In equilibrium, $\delta=0$ and $\lambda > 1$, a bidder with value $x$ bids consistently and repeatedly according to the bidding function $b(x)$
\begin{align*}
 b(x) &= \left\{  \begin{array}{ll}
x & x < X_{(\lambda)} \\
X_{(\lambda)} & x > X_{(\lambda)}
\end{array} \right.
\end{align*}
\end{theorem}

\begin{figure}[tbp]
    \centering
    \begin{minipage}{.49\textwidth}
      \centering
		\includegraphics[height=0.3\textheight]{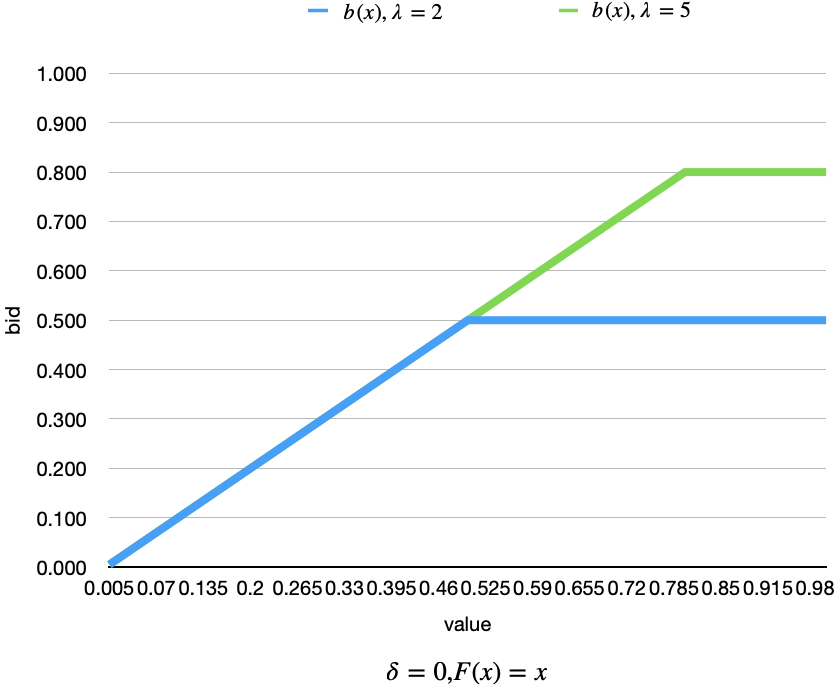}
    \end{minipage}
    \begin{minipage}{.49\textwidth}
      \centering
		\includegraphics[height=0.3\textheight]{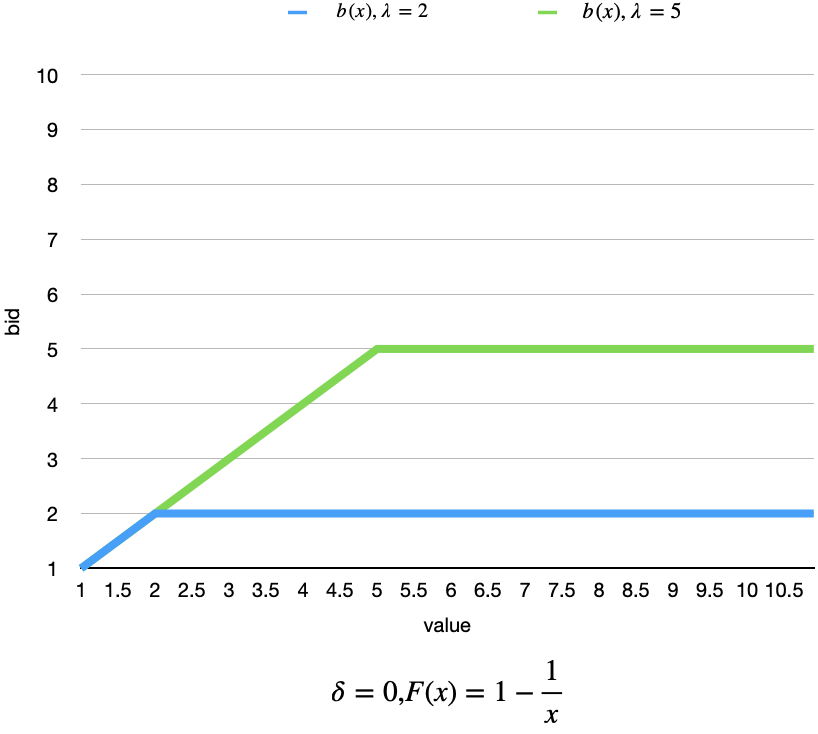}
    \end{minipage}
		\caption{Bidding function with U[0,1]  and $x^2$ power-law distribution}
		\label{bid-no-un}
    \end{figure}

In Figure \ref{bid-no-un} the equilibrium bidding function is plotted for two bidder value distributions: The uniform distribution from $0$ to $1$, and, more realistically, the $x^2$ power-law distribution, with density $1 / x^2$ and c.d.f. $1 - 1 / x$. (According to Pareto (\cite{flux1896vilfredo}), wealth distributions follow a power law with exponent $1 + \alpha$). The values of $\lambda = 2$ and $\lambda = 5$ are used to exemplify both distributions.
\begin{corollary}
\label{price}
In equilibrium, the price paid for an item is $X_{(\lambda)}$.
\end{corollary}

Therefore, price announcements carry no information, and so cannot influence bidder strategy.

According to Proposition \ref{lambda<1}, a high-value bidder, with value $x > X_{(\lambda)}$, will almost surely win a future round, since the Markov chain $(N_t(F(x))_{t\geq0}$ has a stationary distribution. We now evaluate this stationary distribution: What is the distribution of the number of bidders with higher value (and therefore, higher bid) that such a bidder meets on entering the pool?

$N_t(g)$ marks the number of bidders of percentile $> g = F(x)$ at time $t$, where $x > X_{(\lambda)}$. By expectation, every round $\lambda^* = \lambda [1 - g] < \lambda [1 - F(X_{(\lambda)})] = 1$ such new bidders arrive. By Proposition \ref{lambda<1}, the Markov chain $(N_t(g))_{t \geq 0}$ has a unique stationary distribution. $p_n^{(0,\lambda^*)}$ marks  the probability that, in the stationary distribution, $N_t(g) = n$.
Let its probability generating function (PGF) be $$G^{(0,\lambda^*)}(z) := \sum_{n=0}^\infty p_n^{(0,\lambda^*)} z^n$$ 

We use the notation $p_n^{(\delta, \lambda)}$ for the stationary-distribution probability of $n$ bidders in the pool, when the uncertainty is $\delta$ and the arrival mean is $\lambda$. When the context is clear, we abbreviate it to $p_n$. Similarly, we use the notation $G^{(\delta,\lambda)}(z)$ for the PGF, and abbreviate it to $G(z)$ when the context is clear. E.g., the last sentence of the last paragraph can be rephrased: `Let $G(z) = \sum_{n=0}^\infty p_n z^n$ be its probability generating function'.

\begin{proposition}
\label{PGF}
For $\delta = 0$ and $\lambda^* < 1$, the unique stationary distribution has a PGF that satisfies
\begin{align}
\label{stationary}
G(z) = \Big(p_0 + \sum_{n=1}^\infty p_n z^{n-1}\Big) \phi(z)
\end{align}
\noindent whose solution is
\begin{align}
\label{a-pgf}
G(z) = (1 - \lambda^*)  \frac{1 - z}{\phi(z) - z} \phi(z)
\end{align}
\noindent where, for a Poisson arrival process \footnote{The Poisson distribution may be replaced throughout this paper by others, and was chosen for its familiarity.} with mean $\lambda^*$ 
$$\phi(z) := e^{\lambda^* (z - 1)}$$
\end{proposition}

\eqref{a-pgf} is not the PGF of any standard distribution. As it is not a polynomial, the underlying distribution is unbounded. We can derive any element of the distribution using the identity $p_n = \frac{G^{(n)}(0)}{n!}$. For example (for $\lambda^* < 1$), $p_1^{(0, \lambda^*)} = G'(0) = (1 - \lambda^*)(e^{\lambda^*} - 1)$.

The mean of the distribution, i.e., the expectation of the number of entries in the pool with value $> x$, where $\lambda^* = \lambda [1 - F(x)]$, is
$$\E[N_t(F(x))] = \sum_{n=0}^\infty n p_n = G'(1^-) = \frac{G'(1^-)}{G(1^-)} = \frac{1}{z-1} - \frac{\lambda^* \phi(z) - 1}{\phi(z) - z} + \lambda^* \Bigg|_{z \to 1^-} $$
from which, by applying L'H\^{o}pital's rule twice
\begin{proposition}
\label{mean}
When $\delta = 0$ and $\lambda^* = \lambda [1 - F(x)]$ is less than $1$
\begin{align*}
\E[N_t(F(x))]= \frac{\lambda^* (2 - \lambda^*)}{2 (1 - \lambda^*)}
\end{align*}
\end{proposition}

As expected by Proposition \ref{lambda>1}, the mean diverges at $\lambda^*=1$. Proposition \ref{mean} is plotted in {\color{blue} blue} in Figure \ref{mean-no-removal}, alongside the following proposition (in {\color{green} green}).

\begin{proposition}
\label{time-to-win}
Given $\lambda$, the expected number of rounds of a value-$x$ high-value bidder to win a round is $\frac{1}{1 - \lambda [1 - F(x)]}$.
\end{proposition}

Note that $\frac{\lambda^* (2 - \lambda^*)}{2 (1 - \lambda^*)} < \frac{1}{1 - \lambda^*}$ (see Figure \ref{mean-no-removal}), so the mean time to win a round is {\em larger} than the mean pool size, due to the possibility of new bidders arriving.

\subsubsection{An Emergent Posted-Price Mechanism}

Theorems \ref{threshold} and \ref{posted-price} reveal the emergent mechanism that an unending sequential auction with zero uncertainty leads to: A posted-price mechanism, with price  $X_{(\lambda)} = F^{-1}\big(\frac{\lambda - 1}{\lambda}\big)$. The equivalence of the mechanisms is underscored by these facts:
\begin{itemize}
\item Every buyer with value $x$ above the posted price, will almost surely get it for the posted price, after a waiting time averaging $\frac{1}{1 - \lambda [1 - F(x)]}$ rounds (Proposition \ref{time-to-win}).
\item Every buyer with value below the posted price, will almost surely never get it.
\end{itemize}

\section{The Unending Auction with Uncertainty}
\label{uncertainty}

Now we analyze our unending auction model with uncertainty, i.e., with $\delta > 0$: Every round, every bidder in the pool faces an independent event of being removed from the pool, with probability $\delta$.

Recall that in the no-uncertainty case, by Proposition \ref{mean}, the threshold where the pool size became unbounded, and the associated Markov chain had no stationary distribution, was $\lambda=1$. As we shall see, with uncertainty, no matter how small, the expected pool size is {\em always} bounded. It will easily follow that its associated Markov chain has a unique stationary distribution.

We derive an implicit equation for the stationary distribution's (if it exists) PGF, as we did in Proposition \ref{PGF}, this time with $\delta > 0$. Starting at PGF $G(z) := \sum_{n=0}^\infty p_n^{(\delta,\lambda)} z^n = \sum_{n=0}^\infty p_n z^n$, assumed to represent a stationary distribution of the bidder pool size, we get an implicit equation for $G(z)$
\begin{align}
\label{stationary-delta}
G(z) = \Big(p_0 + \sum_{n=1}^\infty p_n [\delta + (1-\delta)z]^{n-1}\Big) \phi(z)
\end{align}

\noindent where $\phi(z) := e^{\lambda(z-1)}$ for a Poisson arrival of bidders with expectation $\lambda$. In transition to \eqref{stationary-delta} from \eqref{stationary}, every factor $z$ was replaced by the PGF of a Bernoulli distribution with probability $1-\delta$, namely $\delta + (1 - \delta)z$. Since, for $p_n$, there are $n-1$ independent Bernoulli events, $z^{n-1}$ is replaced by the PGF of a $\Bin(n-1,1 - \delta)$ distribution, namely $$[\delta + (1 - \delta)z]^{n-1}$$

\begin{figure}[tbp]
    \centering
    \begin{minipage}{.49\textwidth}
      \centering
		\includegraphics[height=0.25\textheight]{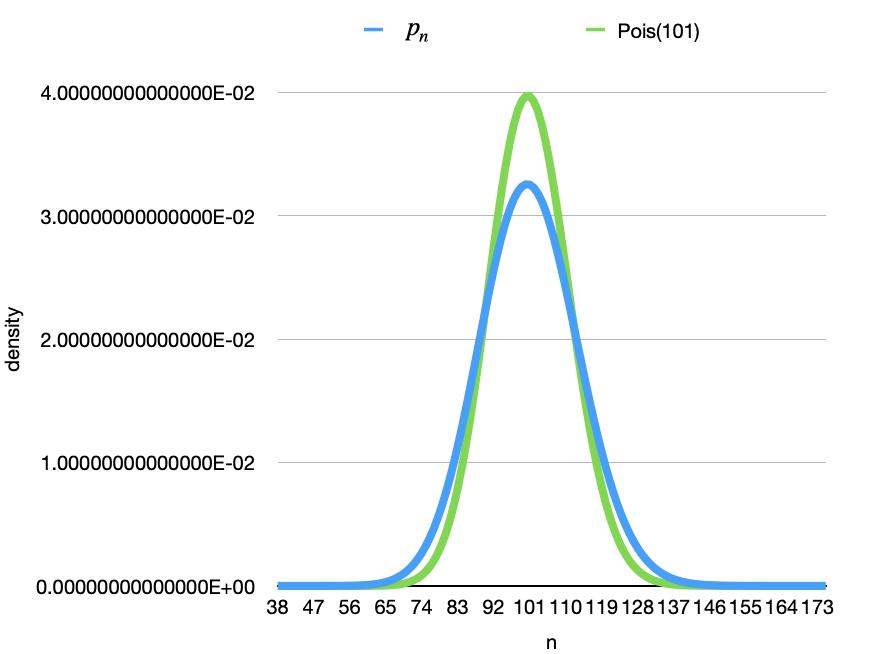}
		\caption{Stationary distribution $\{p_n\}$ with $\lambda=2, \delta=0.01$}
		\label{numeric-p_n}
    \end{minipage}
    \begin{minipage}{.49\textwidth}
      \centering
		\includegraphics[height=0.25\textheight]{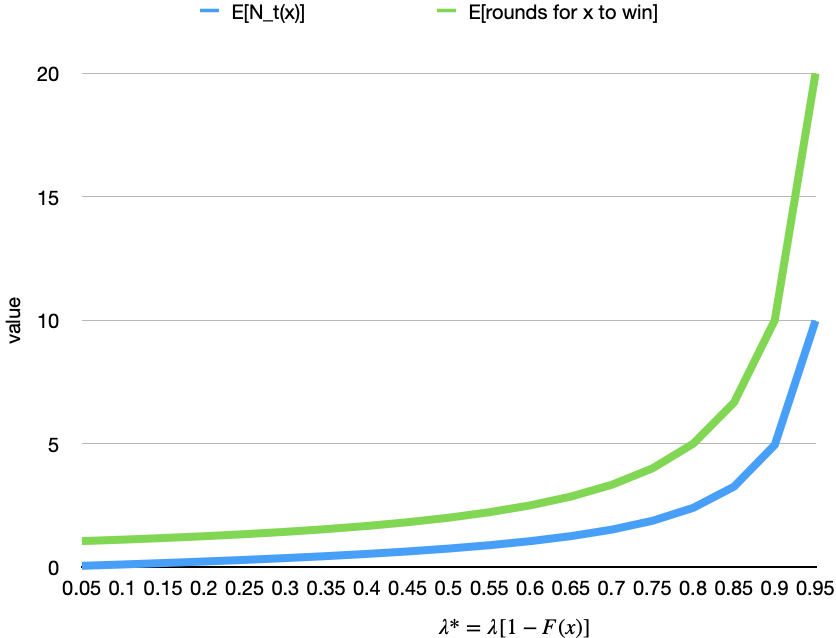}
		\caption{Mean pool size and time to win with values $F^{-1}\big(\frac{\lambda - \lambda^*}{\lambda}\big)$}
		\label{mean-no-removal}
    \end{minipage}
    \end{figure}

An analytic solution of \eqref{stationary-delta} is, unfortunately, not available to us. A numeric solution of \eqref{stationary-delta} for $\lambda=2, \delta=0.01$ is shown in Figure \ref{numeric-p_n} (in {\color{blue} blue}), alongside a Poisson distribution with mean $101$. The distributions seem to have the same mean. The following Proposition confirms this.

\begin{proposition}[Mean Pool Size]
\label{mean-pool-size}
 Let $p_0 \equiv p_0^{\delta, \lambda}$ be the stationary probability for an empty pool.
With bidder uncertainty ($\delta > 0$), the mean size of the bidder pool $N_t$ is finite for any $\lambda$, and equals
$$\E[N_t] = \frac{\lambda - (1-p_0)(1-\delta)}{\delta}$$
\end{proposition}

In the example of Figure \ref{numeric-p_n}, calculation shows $p_0 < 10^{-10}$, so $\E[N_t] \simeq \frac{2 - (1-0.01)}{0.01} = 101$ almost exactly.

\begin{proposition}[Existence of a Stationary Distribution]
\label{exists-statdist}
With bidder uncertainty ($\delta > 0$), any bidder arrival mean $\lambda$, and bidder pool size marked by $N_t$, the Markov chain $(N_t)_{t\geq0}$ over state space $\mathcal{N} := \{N_t = 0, 1, 2, \ldots\}$, has a unique stationary distribution.
\end{proposition}

\subsection{Calculating a Bidding Function}
\label{winner}

Now we start a rather long chain of reasoning, which will lead us to the equilibrium bidding function.

We assume that the auctioneer makes {\em no announcement of prices} paid for units. If the auctioneer does announce prices , such announcements affect equilibrium bidding (unlike in the classical model, and for $\delta > 0$ only). The complexities raised by this may be  intractable, and are left for future work. More details are given in the Appendix. 

We showed in Proposition \ref{exists-statdist} that the bidder pool has a stationary distribution for every $\delta > 0$ and $\lambda$, and gave an implicit equation for it in \eqref{stationary-delta}. Recall Observation \ref{ob}, that every arrival process of bidders with percentile $> g = F(x)$, is a Poisson arrival process with mean $\lambda [1 - g]$. Furthermore, recall Remark \ref{rem}, that we can consider a continuum of bidder pool stationary distributions $\{p_n^{\delta, \lambda[1 -g]}, n = 0, 1, 2, \ldots\}$, each for a different value of $g \in [0,1]$.

This continuum of Markov chains (each representing a partial bidder count) is, simultaneously, each in its own stationary distribution, whenever the general bidder pool (i.e., when $g=0$) is in its stationary distribution. This is because they represent nested subsets of bidders: The bidder pool for parameters $\{\delta, \lambda[1-g]\}$ is a subset of the bidder pool for parameters $\{\delta, \lambda [1 -g']\}$ for every $g' \leq g$, and all are subsets of the general bidder pool.

Now we make another useful observation. 

\begin{observation}
\label{W-ob}
The probability that the round winner has value $\leq x$, equals $p_0^{(\delta, \lambda [1 - F(x)])}$. This is because, by definition, the pool has zero values $> x$ if and only if the winner's value is $\leq x$. 
\end{observation}

Moreover, the above expression shows that this probability, marked $W^{(\delta,\lambda)}(g)$, or $W(g)$ for short, where $g := F(x)$ is value $x$'s percentile, is a function of the {\em percentiles} of the bidders' value distribution $F(x)$.
\begin{definition} [Stationary winner cumulative probability and density]
\label{W-def}
The stationary winner distribution is the distribution of a round winner's value derived from the continuum of all stationary bidder pool distributions $\{p_n^{\delta, \lambda[1-g]}, n=0,1,2, \ldots\}$ for $g \in [0,1]$.

The stationary winner c.d.f. for percentile $g$, given uncertainty $\delta$ and arrival mean $\lambda$, marked $W^{(\delta,\lambda)}(g)$, or $W(g)$ for short, equals  $p_0^{(\delta, \lambda [1 - g])}$. Its density is marked $w^{(\delta,\lambda)}(g)$, or $w(g)$ for short.
\end{definition}

As our model states, we are in search of a stationary bidding function, that, when bid repeatedly and consistently by the bidders, makes a sequential equilibrium. We hypothesize that this is satisfied if the round winners' distribution is stationary. This hypothesis will be justified by showing that it does, indeed, lead to a stationary equilibrium bidding function. 

%The density {\em by value} is not $w(g) = \frac{d}{dg} W(g)$, which is the density by {\em percentile}, but
%$$\frac{d}{dx} W(g) = \frac{d}{dx} W(F(x)) = w(F(x)) f(x) = w(g) f(x)$$

We can now rephrase Theorem \ref{threshold}, which applied to $\delta=0$, in winner distribution terminology:
\begin{align}
\label{Wg}
W(g) = p_0^{(0,\lambda [1 - g])}  &= \left\{  \begin{array}{ll}
1 - \lambda (1 - g) & g > \frac{\lambda-1}{\lambda} \\
0 & g < \frac{\lambda-1}{\lambda}
\end{array} \right.
\end{align}

We can calculate $W(g)$ by solving \eqref{stationary-delta} for all values of parameters $\delta, \lambda[1-g]$, $g \in [0,1]$ (or, practically, for a sample of them), and extracting the values of $p_0^{\delta,\lambda[1-g]}$. One such solution (for $\delta=0.01, \lambda=2, g=0$) was shown in Figure \ref{numeric-p_n}. There, $p_0$ was calculated as $< 10^{-10}$.

It is worth noting that $W(g)$ and $w(g)$ are ``universal'' functions, depending on parameters $\delta$ and $\lambda$, but not on the bidders' value distribution $F(x)$.

\begin{figure}[tbp]
    \centering
    \begin{minipage}{.49\textwidth}
      \centering
		\includegraphics[height=0.3\textheight]{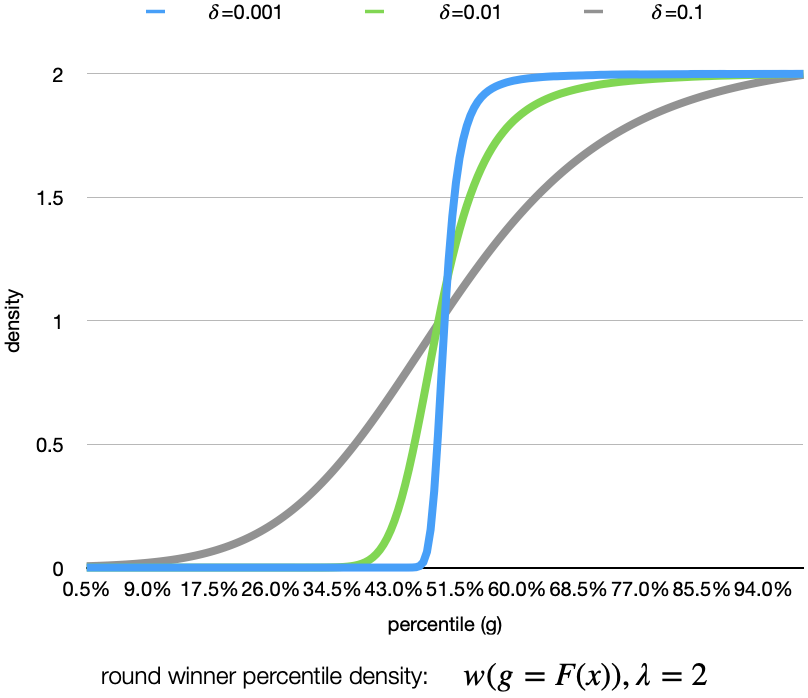}
    \end{minipage}
    \begin{minipage}{.49\textwidth}
      \centering
		\includegraphics[height=0.3\textheight]{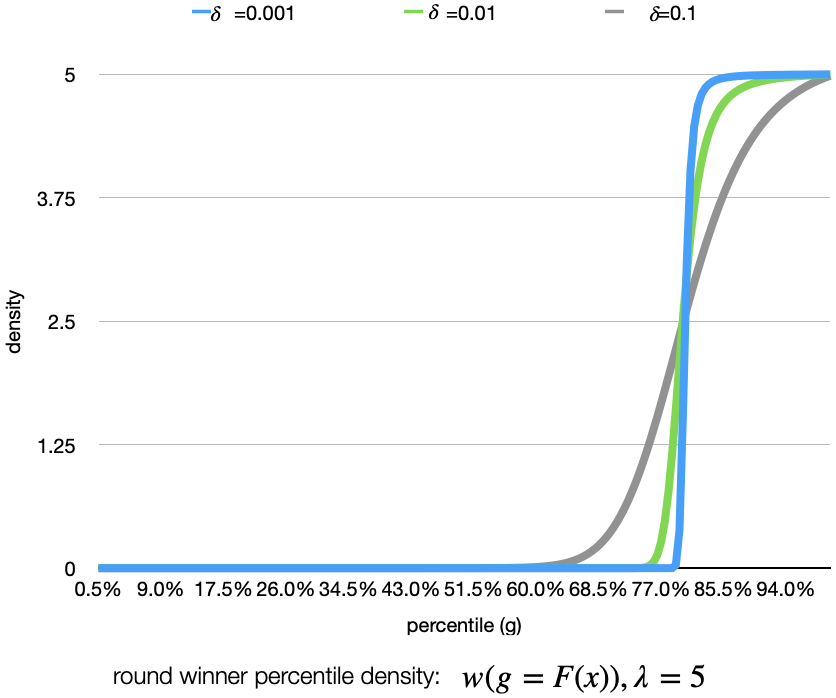}
    \end{minipage}
		\caption{Stationary winner density $w(g)$ by percentile}
		\label{winner-u}
    \end{figure}

$w(g)$ is plotted in Figure \ref{winner-u}, for $\lambda=2$ and $\lambda=5$. Note that $\frac{d}{dx} W(g) = w(g) f(x)$ is the winner density {\em by value}. The plots should be seen in comparison to the 0/1 dichotomy of Theorem \ref{threshold}, which had $\delta=0$ uncertainty. The plots show a ``fuzzy'' dichotomy around the same threshold $X_{(\lambda)}$.

\subsection{Probability of a Bidder's Success}

$W(g)$, introduced in the previous section, is the probability that a $g$-percentile, or lower, bidder will win any particular round. We determine the probability that such a bidder will win an item before being removed, and call this the probability of a bidder's {\em success}.

In the zero-uncertainty unending auction, Propositions \ref{lambda>1} and \ref{lambda<1} provided the value of bidder success: $0$ or $1$, depending on whether $x = F^{-1}(g)$ is below or above $X_{(\lambda)}$, respectively.

Mark this probability by $H(g) \equiv H(F(x))$. It is the probability of winning the first round, or alternatively, not winning the first round, not being removed, then winning the second round, or not winning the first two rounds, again not being removed, then winning the third round, etc. Thus
\begin{align}
\label{H_x}
H(g) = W(g) \big[1 + (1-W(g)) (1-\delta) + (1-W(g))^2 (1-\delta)^2 + \ldots \big] = \frac{W(g)}{1 - (1 - W(g))(1 - \delta)}
\end{align}

\begin{figure}[tbp]
    \centering
    \begin{minipage}{.48\textwidth}
      \centering
		\includegraphics[height=0.3\textheight]{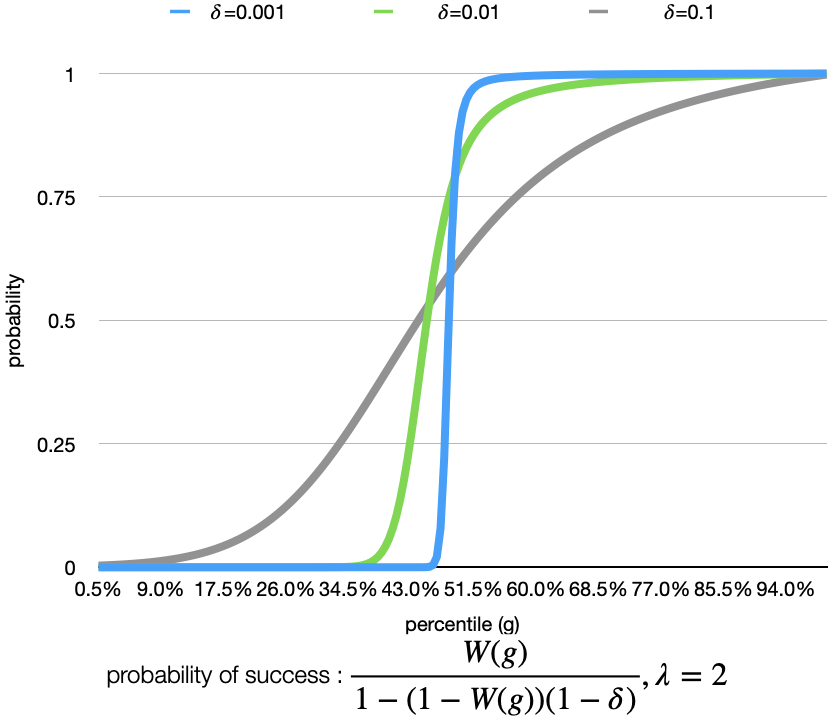}
    \end{minipage}
    \begin{minipage}{.48\textwidth}
      \centering
		\includegraphics[height=0.3\textheight]{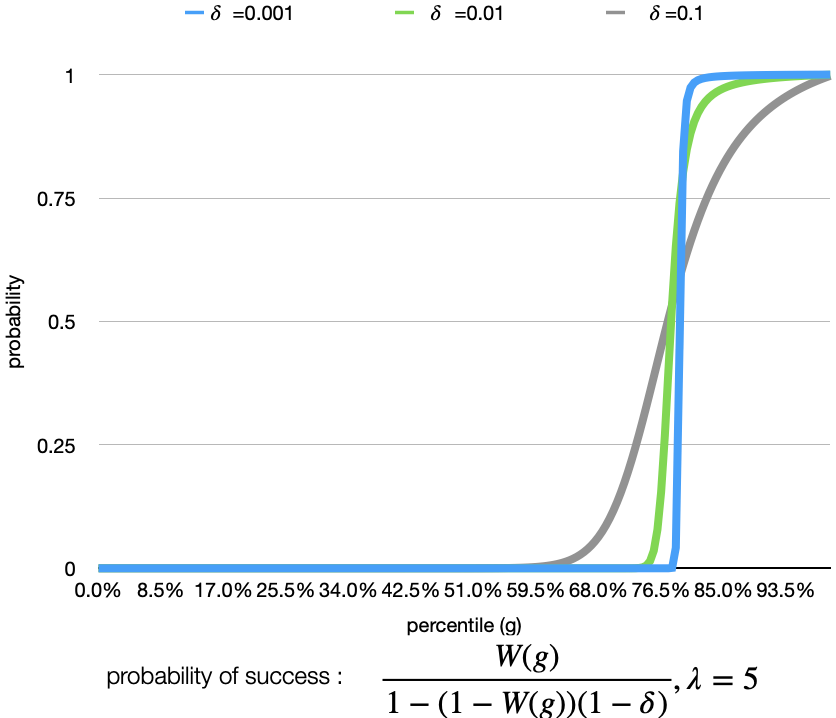}
    \end{minipage}
		\caption{Probability of success $H(g)$ by percentile}
    		\label{probwin}
    \end{figure}

$H(g)$, like $W(g)$, is dependent only on the percentiles of $F(x)$. $H(g)$ is plotted in Figure \ref{probwin} for $\lambda=2$ and $\lambda=5$.

\subsection{The Bidding Function}

 With $\delta = 0$, Theorem \ref{posted-price} showed a simple bidding rule, amounting to a posted-price mechanism, which we were able to deduce with hardly any calculation. With $\delta > 0$, we have no such luxury. Our reasoning is the following.

\begin{enumerate}
\item We are in a stationary distribution. This means every round is, stochastically, the same: The bidder pool size distribution, the round winner distribution (given by $W(g)$), and so is the bidding function $b(x)$, which is the same in the current round,  and every future round.
\item An optimal bidding strategy is in equilibrium with the competing bidders. Only the highest-value bidder should be considered. In the absence of price announcements, its value is an independent random draw from the stationary winner distribution.
\item To be in equilibrium, the bidding strategy must allow no profitable deviations. I.e., in each round, a bidder with value $x$ who decides to bid like a value $y$ bidder $b(y) \neq b(x)$, intending to revert to the equilibrium strategy $b(x)$ in the next round, and having expected revenue for the unending auction marked $V(b(y), b(x), b(x), \ldots)$, maximizes revenue at $y = x$, i.e., 
\begin{align}
\label{dev_one}
\frac{\partial}{\partial y} V(b(y), b(x), b(x), \ldots)\Big|_{y  = x} = 0
\end{align}
Another possible deviation is to impersonate a bidder with value $y \neq x$ for the entire auction, and a condition for this being non-profitable is that 
\begin{align}
\label{dev_all}
\frac{\partial}{\partial y} V(b(y), b(y), b(y), \ldots)\Big|_{y = x} = 0
\end{align}
 \item The expected value from the unending auction of a bidder with value $x$, who follows a $b(x)$ bidding strategy in a first-price auction is
\begin{align}
\label{V_x}
V(b(x), b(x), b(x) \ldots) = H(F(x)) [x - b(x)]
\end{align}
\noindent since $x - b(x)$ is the bidder's gain from bidding $b(x)$ every round, until winning a round, and $H(F(x))$ is the probability that this will happen, given in \eqref{H_x}.
\end{enumerate}

Suppose a bidder with value $x$ deviates to value $y$, for one round only. His expectation is
\begin{align}
\label{V_y}
V(b(y), b(x), b(x), \ldots) = W(F(y)) [x - b(y)] + (1 - \delta) [1 - W(F(y))] H(F(x)) [x - b(x)]
\end{align}
\noindent where the first term of the right-hand side of \eqref{V_y} is the expected revenue in the current round, and the second term is, by \eqref{V_x}, the expected revenue in the remainder of the unending auction, conditioned on {\em not} winning the first round, and not getting removed after the first round.\footnote{\label{g} Our reasoning here is materially different from the standard reasoning in sequential auctions. There, the equilibrium bid makes the bidder indifferent between winning the current round and winning the next round. Here, indifference is impossible, since the bidder's expectation in the the next round is discounted by $1-\delta$. In addition, the current round's highest value is not guaranteed to win the next round, as an arriving bidder may overbid him.}

We apply the first-order condition \eqref{dev_one}, differentiating \eqref{V_y} and substituting $x$ for $y$, obtaining
\begin{align}
\label{first-order}
w(F(x)) f(x) [x - b(x)] - W(F(x)) b'(x) - w(F(x)) f(x) (1 - \delta) H(F(x)) [x - b(x)] = 0
\end{align}

Rearranging \eqref{first-order}, and substituting \eqref{H_x}, we derive

\begin{proposition}[Bidding ODE]
\label{ode-prop}
With bidder uncertainty, $\delta > 0$, stationary winner c.d.f. $W(x)$ and density $w(x)$, derived from the bidder value distribution $F(x)$ through Observation \ref{W-ob}, the bidding function $b(x)$, in equilibrium, satisfies the ordinary differential equation
\begin{align}
\label{ode}
\frac{b'(x)}{x - b(x)} = \frac{w(F(x)) f(x)}{\big[1 + \frac{1-\delta}{\delta} W(F(x)) \big] W(F(x))}
\end{align}
\end{proposition}

The first-order condition for multi-round deviations \eqref{dev_all} leads to the exact same equation \eqref{ode}.

Equation \eqref{ode}, with the boundary conditions $b(0) = 0$, and $b(x) = x$ for $x < \underline{X}$, can be solved, as detailed in the following theorem.

\begin{theorem}[Bidding with Uncertainty]
\label{bidding}
With uncertainty $\delta > 0$,  let the stationary winner c.d.f. be $W(g)$, its density $w(g)$, and the bidder value distribution $F(x)$. Then, in equilibrium, the bidding function $b(x)$ is given by
\begin{align}
\label{bid-function}
b(x) = \Big[\frac{1}{W(F(x))} + \frac{1-\delta}{\delta}\Big] \int_{\underline{X}}^x \frac{z w(F(z)) f(z) }{\big[1 + \frac{1-\delta}{\delta} W(F(z)) \big]^2} dz
\end{align}
\end{theorem}

%It is interesting to derive a bidding function based on percentiles, rather than on values. We can change variables in \eqref{bid-function}, to get a bidding function $b^*(g)$ based on percentile, using the identity $dg = f(x) dx$. The result is dependent on the value distribution $F(x)$  only through the factor $F^{-1}(\gamma)$ in the integrand.
%
%\begin{corollary} [Bidding by Percentile]
%\label{by-percentile}
%\begin{align*}
%b^*(g) := b(F^{-1}(g)) = \Big[\frac{1}{W(g)} + \frac{1-\delta}{\delta}\Big] \int_0^g \frac{F^{-1}(\gamma) w(\gamma)}{\big[1 + \frac{1-\delta}{\delta} W(\gamma) \big]^2} d\gamma
%\end{align*}
%\end{corollary}

\begin{figure}[tbp]
    \centering
    \begin{minipage}{.49\textwidth}
      \centering
		\includegraphics[height=0.29\textheight]{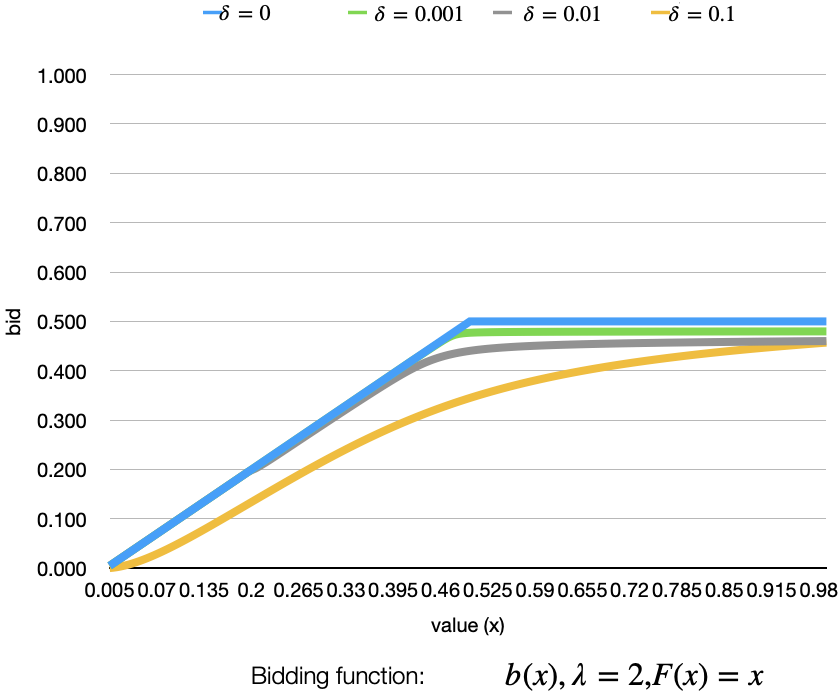}
    \end{minipage}
    \begin{minipage}{.49\textwidth}
      \centering
		\includegraphics[height=0.29\textheight]{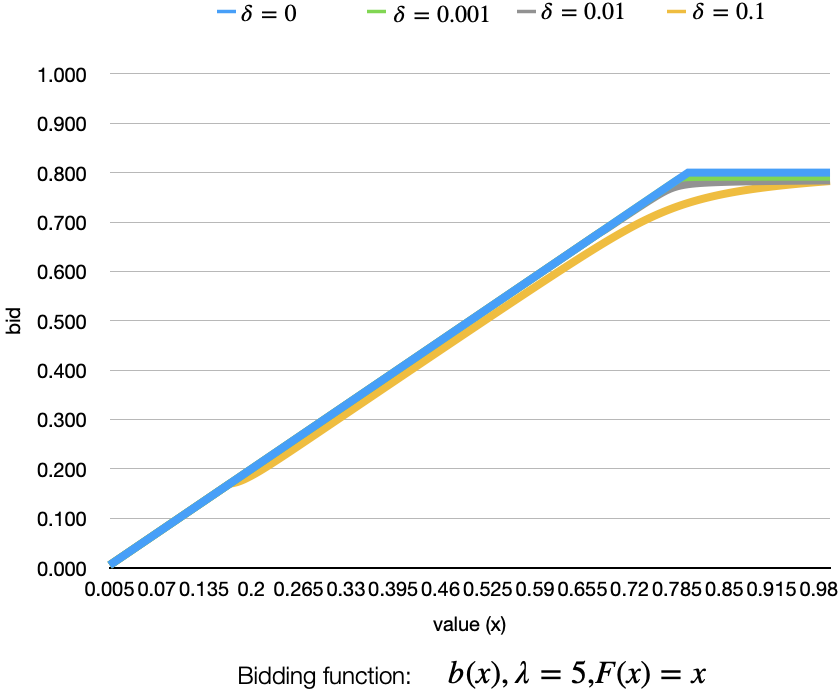}
    \end{minipage}
		\caption{Bidding function: $b(x)$ for $U(0,1)$ distribution}
    		\label{bid-u}
    \end{figure}

\begin{figure}[tbp]
    \centering
    \begin{minipage}{.49\textwidth}
      \centering
		\includegraphics[height=0.3\textheight]{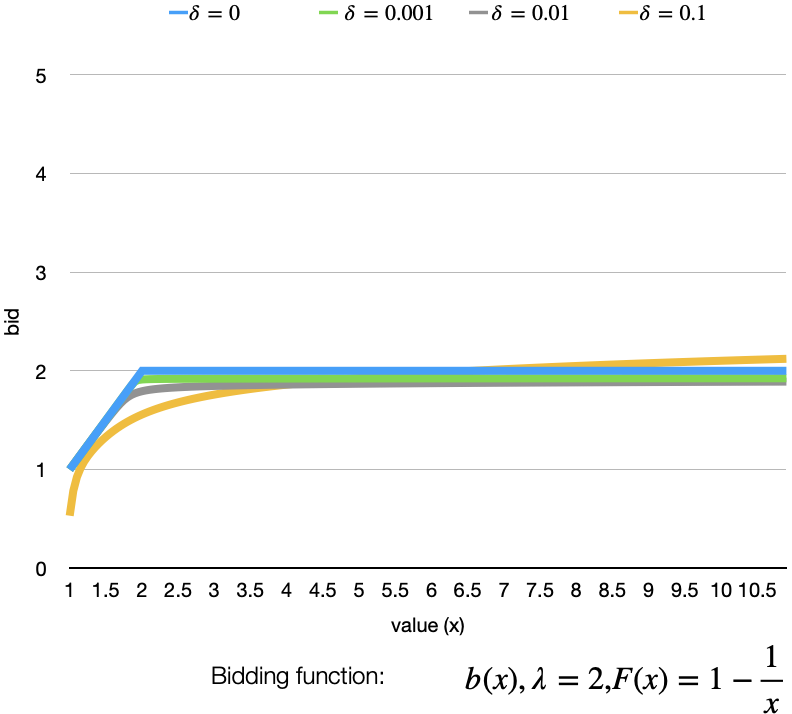}
    \end{minipage}
    \begin{minipage}{.49\textwidth}
      \centering
		\includegraphics[height=0.3\textheight]{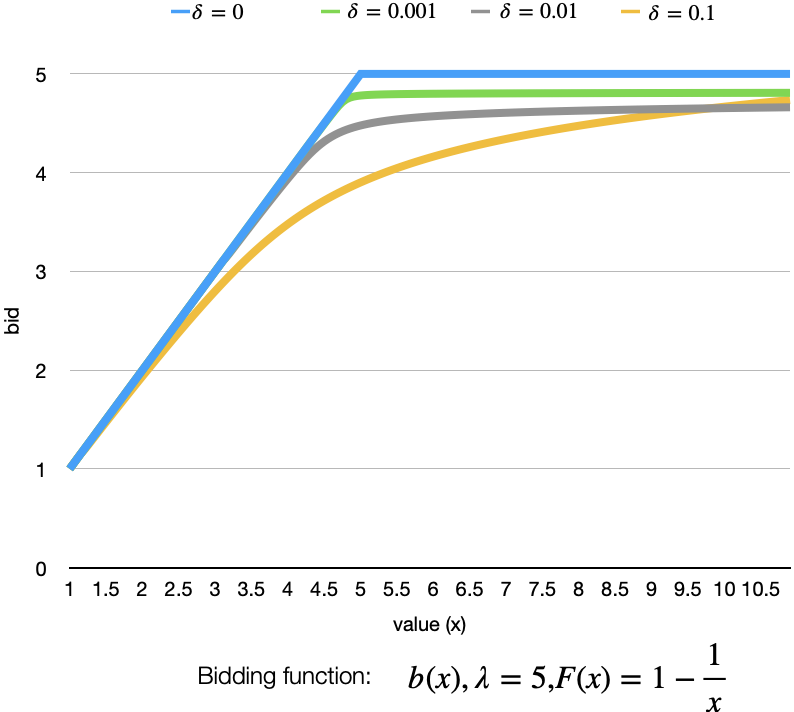}
    \end{minipage}
		\caption{Bidding function: $b(x)$ for $x^2$ power-law distribution}
    		\label{bid-p}
    \end{figure}

$b(x)$ is plotted in Figures \ref{bid-u} and \ref{bid-p} for two value distributions, each for two values of $\lambda$.

\subsection{Bidder Expectation}
\label{expectation}

\begin{figure}[bp]
    \centering
    \begin{minipage}{.49\textwidth}
      \centering
		\includegraphics[height=0.3\textheight]{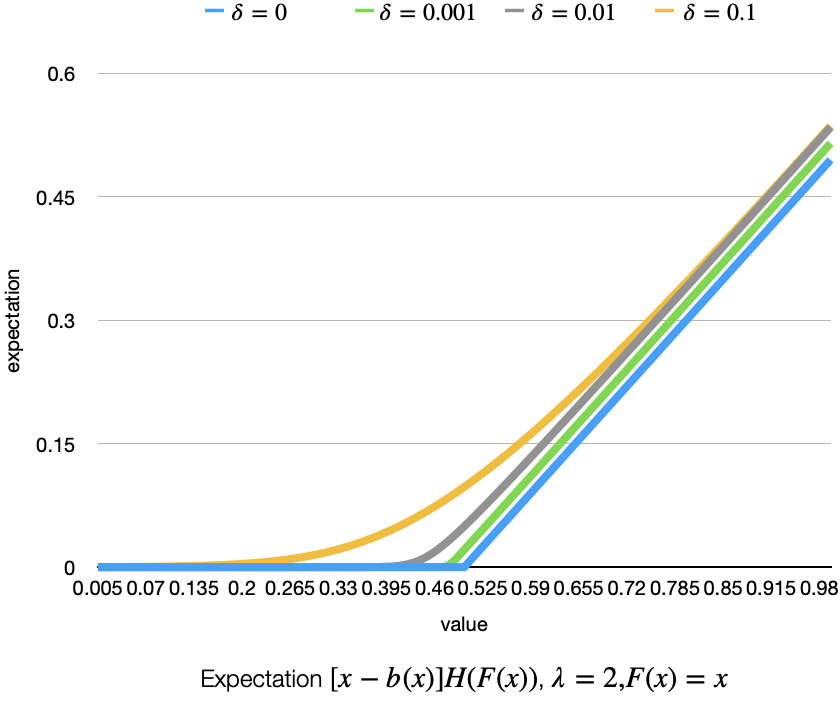}
    \end{minipage}
    \begin{minipage}{.49\textwidth}
      \centering
		\includegraphics[height=0.3\textheight]{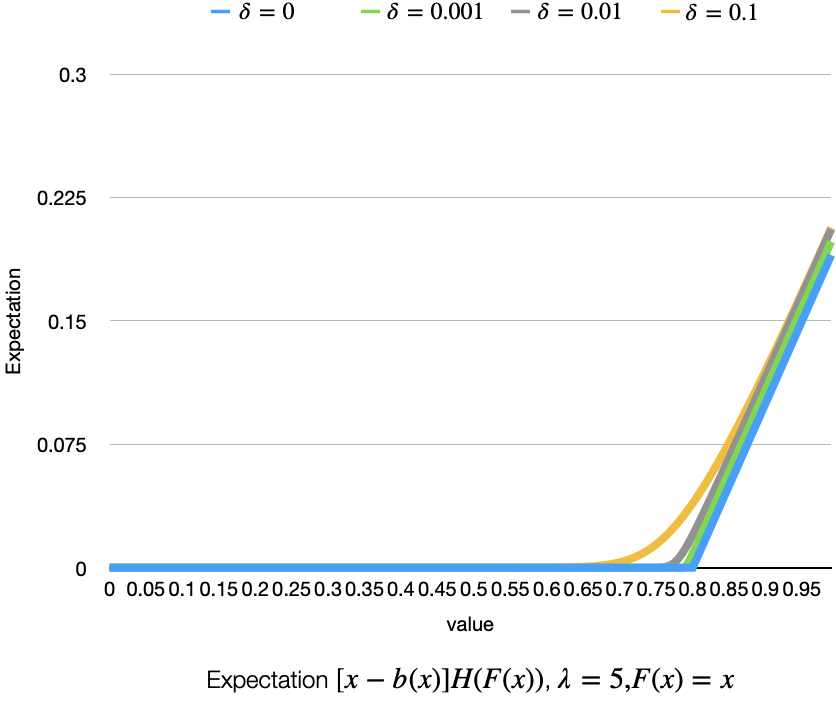}
    \end{minipage}
		\caption{Bidder expectation for $U(0,1)$ distribution}
    		\label{expect-u}
    \end{figure}

\begin{figure}[bp]
    \centering
    \begin{minipage}{.49\textwidth}
      \centering
		\includegraphics[height=0.3\textheight]{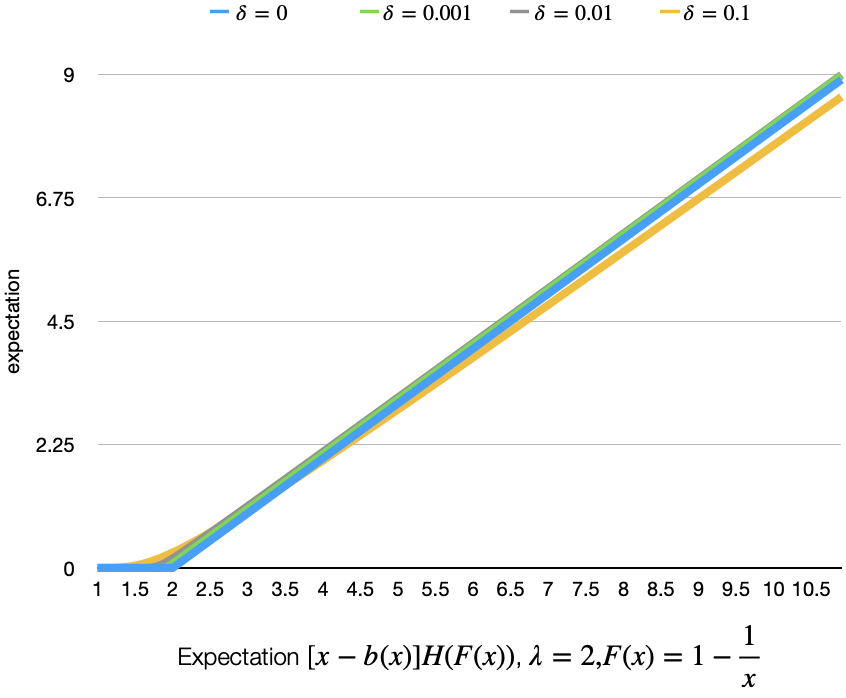}
    \end{minipage}
    \begin{minipage}{.49\textwidth}
      \centering
		\includegraphics[height=0.3\textheight]{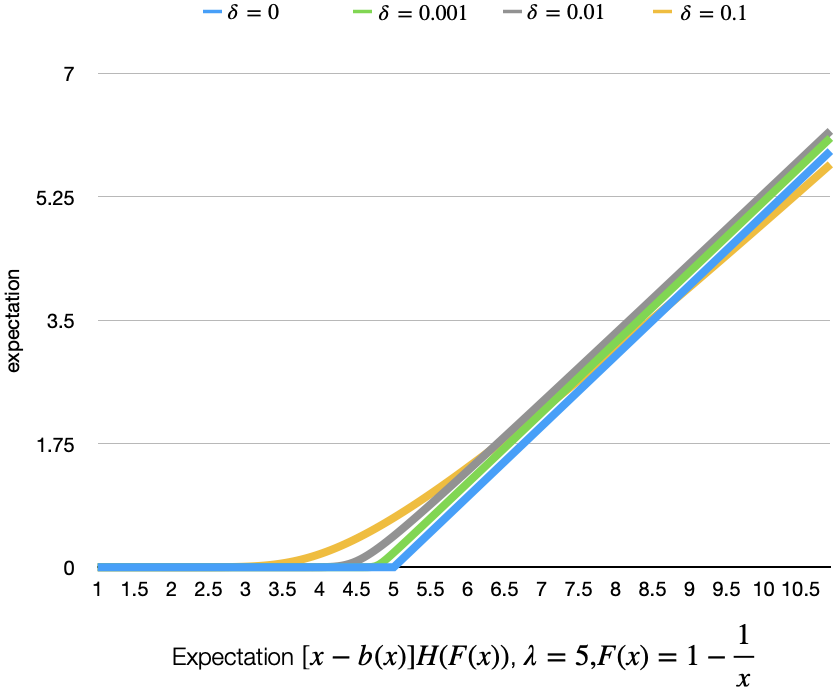}
    \end{minipage}
		\caption{Bidder expectation for $x^2$ power-law distribution}
    		\label{expect-p}
    \end{figure}

As noted in \eqref{V_x}, a bidder with value $x$ has expectation $[x - b(x)] H(F(x))$. The expectation is plotted in Figures \ref{expect-u} and \ref{expect-p}, for two value distributions, each for two values of $\lambda$.

\subsection{Is Uncertainty Good or Bad for Bidders?}

Common sense suggests that uncertainty $\delta > 0$ is bad for bidders. A risk of being summarily expelled from the auction is not the sort of feature that leads to gain. Nevertheless, uncertainty turns out to be beneficial to low-value bidders (value below $X_{(\lambda)}$), and even for high-value bidders (value above $X_{(\lambda)}$), whose values are not `too high'.

Moreover, this applies not only to the uncertainty model we defined in Section \ref{model}, but to  {\em any} model of uncertainty, comprising any value-discounting schedule, and/or deterministic or probabilistic lifetime limiting scheme. Here is the definition of the models covered:
 
\begin{definition}[A General Uncertainty Model]\footnote{Almost all schemes covered by Definition \ref{uncertain} lead to a heterogeneous bidder population (i.e., not drawn from a single distribution) due to differentiation by age. In such auctions, the simplifying assumption that the highest-value bidder wins in equilibrium is wrong. As a result, the analysis of such auctions is probably intractable. We are not aware of any heterogenous-population auction successfully solved analytically and published. (\cite{kittsteiner2004declining}, for example, adopted a discounting method that retains a homogeneous population, albeit not in our unending setting). Our adopted removal process is a homogeneous exception, and that is its only justification.
}
\label{uncertain}

\begin{itemize}
\item Fix $\delta \in (0,1]$. For every bidder $i$ and every round $m$
\begin{itemize}
\item Bidder $i$ has value $v_i^m$ at round $m$.
\item A bidder with value $v_i^m = 0$ is removed from the pool.
\item $v_i^{m+1} = v_i^m r_i^m(\delta)$.
\item $r_i^m(\delta) = R_i^m(\delta)$, where $R_i^m(\delta) \in [0,1]$, with probability $Q_i ^m(\delta)$, and $0$ otherwise, and
\item $Q_i^m(0) = 1, R_i^m(0) = 1$ and 
\item $Q_i^m(\delta), R_i^m(\delta)$ are continuous in $\delta$.
\end{itemize}
\end{itemize}
In other words, a possibly individual, possibly stochastic discounting/removal per-round scheme with a real parameter $\delta$ that does not discount/remove at all when $\delta=0$.
\end{definition}

To support this claim, we prove a general lemma.
\begin{lemma}
\label{increasing}
Let $A(x;\delta)$ be some real function of $x, \delta \in \mathbb{R}_{\geq 0}$. Assume $A(x; \delta)$ continuous everywhere in both its arguments, except possibly at $\delta=0$. Furthermore, assume that for every $x \in [0, X]$ and for every $\delta > 0$, $A(x;0) \leq A(x;\delta)$. Then, there exists $X^{+} \geq X$ such that for every $x \in [0, X^{+})$ there exists $\delta_x > 0$ such that $A(x, \delta)$ is non-decreasing in $[0, \delta_x]$.
\end{lemma}

With the help of this Lemma, we prove two favorability-of-uncertainty results, the first regarding equilibrium bidding, the second regarding bidder expectation in equilibrium. Both apply to any uncertainty model defined in Definition \ref{uncertain}.

The generality of these results stem from two simple facts: The results in the baseline case, the zero-uncertainty auction, presented in Section \ref{markov}, are the worst possible ones for low-value bidders. No model can do worse. This, and the continuity by value, and by $\delta$, carry the low-value bidders' gain into high-value bidders' territory. As plots for example distributions show, this is not a negligible effect, and it might affect most, even all bidders.

A less technical explanation is this: Low-value bidders bid their value with no uncertainty, as they have zero chance of winning a round. Uncertainty gives them some probability of winning, and to capitalize on that they must bid {\em lower}. As for high-value bidders, we found they all bid the threshold value with no uncertainty. Why this value? Because it is the lowest that outbids all low-value bidders. With uncertainty, as low-value bidders bid lower, high-value bidders can outbid all with a lower bid. But this is only one side of the equation: Uncertainty means that their surplus (value - bid) is at risk. The risk can be mitigated by bidding higher, so as to win faster. This is why we see high-value bidders bidding (and paying) less, but only while their surplus is not ``too high''.

The favorability results are due to at least some low-value bidders experiencing uncertainty. When none do, as in, for example, Section \ref{a-few} below, no one gains from uncertainty.

\begin{theorem}[Bids Decreasing by Uncertainty]
\label{bid-less}
Let $b(x | \lambda, \delta)$ be the bidding function in equilibrium for arrival and uncertainty process parameters $\lambda$ and $\delta$, respectively. Let $X_{(\lambda)}$ be the threshold for which $F(X_{(\lambda)}) = \frac{\lambda-1}{\lambda}$.
Then, subject to any general discounting model, there exists $\delta^* > 0$ and $X^* \geq X_{(\lambda)}$ such that 
\begin{enumerate}
 \item $b(x | \lambda, \delta) \leq b(x | \lambda, 0)$ for every $\lambda, \delta \leq \delta^*$ and $x \leq X^*$. \label{i}
 \item $b(x | \lambda, \delta)$ is non-increasing in $\delta$ for every sufficiently small $\delta$. \label{ii}
 \item $b(x | \lambda, 0) - b(x | \lambda, \delta)$ is maximal at $x = X_{(\lambda)}$. \label{iii}
\end{enumerate}
\end{theorem}

\begin{theorem}[Bidder Expectation Increasing by Uncertainty]
\label{expect-more}
Mark bidder expectation by $Z(x | \lambda, \delta) := Z(x) = [x - b(x)] H(F(x))$, by \eqref{V_x}.
Then, subject to any general discounting model, there exists $\delta^* > 0$ and $X^* \geq X_{(\lambda)}$ such that 
\begin{enumerate}
\item $Z(x | \lambda, \delta) \geq Z(x | \lambda, 0)$ for every $\lambda, \delta \leq \delta^*$ and $x \leq X^*$, and
\item $Z(x | \lambda, \delta)$ is non-decreasing in $\delta$ for every sufficiently small $\delta$.
\end{enumerate}
\end{theorem}

\section{Discussion}
\label{discussion}

\subsection{Multiple Winners}
Generalizing our model to multiple winners while retaining the main posted-price mechanism result is straightforward, and is described in the Appendix. In a nutshell, with $\mu$ winners per round, most results hold with $\lambda$ replaced by $\frac{\lambda}{\mu}$.

\subsection{A Few Impatient Bidders}
\label{a-few}

What happens when a few impatient ($\delta > 0$) high-value bidders are embedded in a sea of patient ($\delta=0$) bidders?

The mass of patient bidders will still enforce a posted-price mechanism of $X_{(\lambda / \mu)}$ (in a multi-unit auction), due to their numbers. The few impatient ones will not be able to capitalize on decreasing bids by low-value bidders, and will be forced to bid {\em more} than $X_{(\lambda / \mu)}$, and progressively more according to their level of impatience, and the private value they wish to protect.

This is probably what we are seeing in the Bitcoin mempool snapshot of Figure \ref{mempool}.

\subsection{A Change in Bidder Arrival Rates}

So far we assumed that the arrival process of new bidders has a fixed mean $\lambda$.
What happens when $\lambda$ changes to $\lambda^*$, due to a shock, or due to a planned event?

The system will transition to a new dichotomy between low-value bidders and high-value bidders, now separated by $X_{(\lambda^*)}$. The bidder pool size will settle around $O(\lambda^* / \delta)$.

But, the transition dynamics differ according to whether $\lambda^* > \lambda$ or $\lambda^* < \lambda$.

Assume all bidders are able to calculate and follow a new equilibrium instantaneously.

If $\lambda^* > \lambda$, prices will rise to $X_{(\lambda^*)}$, and this will happen fast, as new high-value bidders (with value above $X_{(\lambda^*)}$), will shut out bidders with value between $X_{(\lambda)}$ and $X_{(\lambda^*)}$, who have suddenly become low-value bidders.

If $\lambda^* < \lambda$, the transition will not end until the unblocked `congestion' of bidders with value between $X_{(\lambda^*)}$ and $X_{(\lambda)}$, who have been `stuck' in the bidder pool, and have suddenly become high-value bidders, is cleared. Without uncertainty, this will merely mean a very long waiting time for such bidders. With uncertainty, the `congested' bidders may not afford the risk of a long wait, and, will compensate by higher bidding and prices. Thus we expect the drop in prices to be gradual.

\subsection{Unending Second Price Auctions}
\label{second-price}

The Classical approach to solving second price sequential auctions for equilibrium, according to (\cite{krishna2009auction} and others, is to use the Revenue Equivalence Theorem to derive second price auction results from first price auction results. This should hold in the unending setting as well, so, second price auction bidding functions may be derived by demanding that their revenue be the same as for first price auctions.

Another insight, which leads to a different method, is to observe that the Markov chains we derived hold equally well for all efficient auctions (i.e., those where the highest value wins). The derivations of the bidding functions would diverge only at \eqref{V_y}, the formulation of the first-order condition.

A second price auction, or any other efficient auction, is unlikely to have a qualitative effect on results. Moreover, clearly the no-uncertainty result for Second Price auctions is exactly the same posted-price emulation derived for First Price auctions.

\bibliographystyle{ACM-Reference-Format}
\bibliography{infomarkets}

\newpage
\appendix

\section {Proof of Proposition \ref{lambda>1}}

\begin{proof}
Since $\lambda - 1 < 0$, $\lim \inf S_t = -\infty$. It follows from \eqref{N_t} that $\lim \inf A_t/t = 1 - \lambda$. So the chain is positively recurrent (i.e., from any starting point, $N_t = 0$ will occur a.s., in bounded time). Since the Poisson distribution ascribes a positive mass to any non-negative integer, the chain is irreducible and aperiodic. By a standard result of Markov theory, the chain has a unique stationary distribution (see \cite{lalley2016markov} Theorem 19).
\qed
\end{proof}

\section {Proof of Proposition \ref{lambda=1}}

\begin{proof}
Since the Poisson distribution is unbounded, $\lim \sup S_t = -\infty$. By \eqref{N_t} $\lim \sup A_t = \infty$, and thus $\lim A_t = \infty$. In other words, the future occurrence $N_t = 0$ is assured. On the other hand, its probability is $\lim\inf S_t = 1 - \lambda = 0$. In other words, the expected time to $N_t = 0$ is unbounded.
\qed
\end{proof}

\section {Proof of Theorem \ref{threshold}}

\begin{proof}
With $\lambda > 1$, a high-value bidder joins an arrival sub-stream with mean $\lambda [1 - F(x)]$ of bidders with value at least $x$, where $\lambda [1 - F(x)] \leq \lambda [1 - F(X_{(\lambda)})] = 1$. So value $x$ is guaranteed, by Proposition \ref{lambda<1}, to become the largest value in the pool at some future round, and so be a winner. On the other hand, a low-value bidder joins an arrival sub-stream with mean $\lambda [1 - F(y)]$, of bidders with value at least $y$, where $\lambda [1 - F(y)] > \lambda [1 - F(X_{(\lambda)})] = 1$. So this bidder, by Proposition \ref{lambda>1}, joins a sub-pool of values above $y$ that almost surely will never be empty. In other words, value $y$ will always have a higher competitor, and will almost surely never win a round.
\qed
\end{proof}

\section {Proof of Theorem \ref{posted-price}}

\begin{proof}
Low-value bidders almost surely cannot win a round. Nevertheless, a strategy where they do not bid their values cannot be an equilibrium. They cannot bid more, due to individual rationality, and a lower bid will cause high-value bidders to deviate and bid lower, in turn tempting the low-value bidders to overbid them.

It follows that high-value bidders, who expect to win a round by Theorem \ref{threshold}, must bid more than all low-value bidders, and so at least $X_{(\lambda)}$, to win. This bid is sufficient, since their  competitors for winning are counted by a Markov chain with $g > \frac{\lambda-1}{\lambda}$, which will visit state $N_t(g) = 0$ almost surely.
\qed
\end{proof}

\section{Proof of Proposition \ref{PGF}}

\begin{proof}
We evaluate $G(z)$, given that it is stationary. At the start of round $t+1$, a  winner is removed, unless $n=0$, so $G(z)$ is transformed into $p_0 + \sum\limits_{n=1}^\infty p_n z^{n-1}$. Then a $\lambda^*$-Poisson arrival of new bidders occur, independently of all previous occurrences. The PGFs of independent events are multiplied to form the resulting PGF of the resulting event. So $G(z)$ is transformed further into $\Big(p_0 + \sum\limits_{n=1}^\infty p_n z^{n-1}\Big) \phi(z)$.

Now, if $G(z)$ is a stationary distribution, this resulting PGF is, by definition, equal to the original PGF. In other words
\begin{align*}
G(z) = \Big(p_0 + \sum\limits_{n=1}^\infty p_n z^{n-1}\Big) \phi(z)
\end{align*}

So
\begin{align*}
G(z) = \Big(p_0 + \frac{G(z) - p_0}{z}\Big) \phi(z)
\end{align*}

\noindent leading to
\begin{align*}
G(z) = p_0  \frac{1 - z}{\phi(z) - z} \phi(z)
\end{align*}

Let us evaluate the above for $1^-$. Every PGF satisfies $G(1^-) = 1$. By L'H\^{o}pital's rule, $1 = p_0 \frac{1}{1 - \lambda^*}$, and so
\begin{align}
\label{p0}
p_0 = 1 - \lambda^* = 1 - \lambda [1 - F(x)]
\end{align}

(We already knew this last equality from Proposition \ref{lambda<1}.) So the pool's stationary probability is determined by its PGF
\begin{align*}
G(z) = (1 - \lambda^*)  \frac{1 - z}{\phi(z) - z} \phi(z)
\end{align*}

As claimed.
\end{proof}

\section {Proof of Proposition \ref{time-to-win}}

\begin{proof}
The bidder must wait for the event $N_t(F(x)) = 0$ to win. In the stationary distribution, each round, the probability for this is $p_0 = 1 - \lambda^* = 1 - \lambda [1 - F(x)]$. The number of rounds the bidder must wait for such an event has expectation $1 / p_0$ (see \cite{lalley2016markov} Corollary 12), as claimed.
\qed
\end{proof}

\section{Proof of Proposition \ref{mean-pool-size}}

\begin{proof}
Note that $\E[N_t] = \sum\limits_{n=0}^\infty n p_n = G'(1)$. Differentiating \eqref{stationary-delta}
\begin{align*}
G'(z) &= \lambda G(z) +  (1 - \delta) \Big(\sum\limits_{n=2}^\infty (n - 1) p_n [\delta + (1-\delta)z]^{n-2}\Big) \phi(z)
\end{align*}

Substituting $z=1$
\begin{align*}
G'(1) &= \lambda + (1 - \delta) \sum\limits_{n=2}^\infty (n - 1) p_n \\
&= \lambda + (1 - \delta) \Big(\sum\limits_{n=2}^\infty n p_n - \sum\limits_{n=2}^\infty  p_n\Big) \\
&= \lambda + (1 - \delta) \Big((G'(1) - p_1) - (1 - p_0 - p_1)\Big) \\
&= \lambda + (1 - \delta) \Big(G'(1) - (1 - p_0)\Big)
\end{align*}
So
\begin{align*}
\delta G'(1) &= \lambda - (1 - \delta)(1- p_0) \\
\E[N_t] = G'(1) &=  \frac{\lambda - (1-p_0)(1-\delta)}{\delta}
\end{align*}
\end{proof}

\section{Proof of Proposition \ref{exists-statdist}}

\begin{proof}
We need only prove that any of the Markov states has a positive probability, i.e., that it is positively recurrent, from which (since the unbounded support of the Poisson distribution guarantees that the Markov chain is irreducible and aperiodic) it follows that there exists a unique stationary distribution (see \cite{lalley2016markov} Theorem 19).

For $\lambda < 1$, all states are positively recurrent even when $\delta=0$, by Proposition \ref{lambda<1}, so {\em a fortiori} for $\delta>0$.

For $\lambda \geq 1$, by Proposition \ref{mean-pool-size} the mean pool size is finite and strictly positive, since $(1-p_0)(1-\delta) < 1$. It follows that at least one of the states has positive probability.
\qed
\end{proof}

\section{Proof of Theorem \ref{bidding}}

\begin{proof}
Differentiate \eqref{bid-function} to derive $b'(x)$. 
\begin{align*}
b'(x) &= -\frac{w(F(x)) f(x)}{W(F(x))^2} \int\limits_{\underline{X}}^x \frac{z w(F(z)) f(z) }{\big[1 + \frac{1-\delta}{\delta} W(F(z)) \big]^2} dz + \Big[\frac{1}{W(F(x))} + \frac{1-\delta}{\delta}\Big] \frac{x w(F(x)) f(x) }{\big[1 + \frac{1-\delta}{\delta} W(F(x)) \big]^2} \\
&= -\frac{w(F(x)) f(x)}{W(F(x))^2} \frac{b(x)}{\frac{1}{W(F(x))} + \frac{1-\delta}{\delta}} + \frac{x w(F(x)) f(x) }{W(F(x)) \big[1 + \frac{1-\delta}{\delta} W(F(x)) \big]} \\
&= \frac{w(F(x)) f(x) }{\big[1 + \frac{1-\delta}{\delta} W(F(x)) \big] W(F(x))} [x - b(x)]
\end{align*}

Thus
\begin{align*}
\frac{b'(x)}{x - b(x)} = \frac{w(F(x)) f(x) }{\big[1 + \frac{1-\delta}{\delta} W(F(x)) \big] W(F(x))}
\end{align*}
\noindent which is identical to \eqref{ode}.
\end{proof}

\section{Proof of Lemma \ref{increasing}}

\begin{proof}
Assume not, and there is some $x_D$ for which this is not true. Let $\delta_D$ be the smallest value of $\delta$ which violates this, i.e., $\delta_D := \lim\inf \{\delta_1 \geq 0 | \exists \delta_2 > \delta_1 \land A(x_D, \delta_1) \geq A(x_D, \delta_2)\}$. If $\delta_D > 0$, then this apparent counterexample is eliminated by setting $\delta_x = \delta_D$.

Otherwise, $\delta_D = 0$. If $x \leq X$ we have a contradiction. This leaves the case $x > X$. Let $x_H$ be the $\lim\inf$ of $x$ for which this case exists. If $x_H = X$ we again have a contradiction. Otherwise, all exceptions to the lemma can be eliminated by setting $X^{+} = x_H$.
\end{proof}

\section{Proof of Lemma \ref{bid-less}}

\begin{proof}
The bidding function, by its definition in \eqref{bid-function}, is continuous in both $x$ and $\delta$, except at $\delta=0$. By individual rationality, $x - b(x) \geq 0$. Defining $A(x;\delta) := -b(x | \lambda, \delta)$, item \ref{i} follows directly from the above, item \ref{ii} is established by Lemma \ref{increasing}, and item \ref{iii} is a consequence of  the fact that $b'(x) \leq 1$, as a bid gradient exceeding $1$ cannot form an equilibrium; a bidder at the steepest point of such a gradient would always have an incentive to bid lower.
\qed
\end{proof}

\section{Proof of Lemma \ref{expect-more}}

\begin{proof}
For $\delta=0$ low-value bidders have zero expectation, while for $\delta > 0$, their expectation is non-negative. Setting $A(x;\delta) := Z(x | \lambda, \delta)$, the Theorem follows from Lemma \ref{increasing} for every $\lambda$.
\qed
\end{proof}

\section{Multiple Winners Model and Results}
\label{multiple}

Let us change our model of Section \ref{model} in one respect and one respect only: Every round, the $\mu$ highest bids (instead of the single highest one) win an item, each paying their own bid, and leaving the bidder pool.

\begin{theorem}[Bidding with Multiple Winners]
\label{posted-price-mu}
In an unbounded pool and $\delta = 0$, bidders with value above $X_{(\lambda/\mu)} := F^{-1}(\frac{\lambda - \mu}{\lambda})$ will almost surely win a round, while bidders below $X_{(\lambda/\mu)}$ will almost surely never win a round.
\end{theorem}

\begin{proof}
Let $x$ be the value of a bidder. Let $N_t$ be the number of bidders in the pool at round $t$ {\em whose value is $> x$}, and let $\Lambda_t$ be the number of new bidders arriving at round $t$, {\em who have value $> x$}. Note that $\Lambda_t$ is a subset of a Poisson arrival process with expectation $\lambda$, and it has itself a Poisson distribution with expectation $\lambda^* := \lambda [1 - F(x)]$. Then $(N_t)_{t \geq 0}$ is a Markov chain defined by the relation 
\begin{align}
\label{relation-mu}
N_{t+1} = (N_t - \mu)_+ + \Lambda_t
\end{align}

Note that the bidder with value $x$ will win a round whenever $N_t < \mu$. In other words, if the chain is recurrent. We show that this will happen almost surely when $x > X_{(\lambda / \mu)}$, and will never happen almost surely when $x < X_{(\lambda / \mu)}$. In other words, if the chain is transient.

Define $S_t = \sum\limits_{\tau=1}^t (\Lambda_t - \mu)$ and $A_t = \sum\limits_{\tau=1}^t \sum\limits_{n = 0}^{\mu-1} \mathbbm{1}_{\{N_t \leq n\}}$. Then 
\begin{align}
\label{N_t_mu}
N_t = N_0 + S_t + A_t
\end{align}

By the Law of Large Numbers $\lim \inf S_t / t = \lambda^* - \mu$, and so
\begin{itemize}
\item If $\lambda^* > \mu$, $N_t$ grows without limit: $\lim \sup N_t = +\infty$. The chain is transient, i.e., there is a positive probability that, starting from $N_t = k$, $N_T \geq \mu$ for every $T > t$, and this probability limits at $1$ as $k \to \infty$.

\item If $\lambda^* \leq \mu$, since $\lambda^* - \mu \leq 0$, $\lim \inf S_t = -\infty$. It follows from \eqref{N_t_mu} that $\lim \inf A_t/t = \mu - \lambda^*$. So the chain is recurrent. Since the Poisson distribution ascribes a positive mass to any non-negative integer, the chain is irreducible and aperiodic. 
\end{itemize}
\end{proof}

From this, using the same logic of Section \ref{markov}, replacing $X_{(\lambda)}$ by $X_{(\lambda / \mu)}$, parallel results to those of Section \ref{markov} follow.

\begin{corollary}[Multiple Winners Markov Chain]
In an unbounded pool with zero uncertainty, with $\mu < \lambda$ winners per round, the number of bidders with value $> x$, where $x \geq X_{(\lambda / \mu)}$, constitute a recurrent Markov with a unique stationary distribution.
\end{corollary}

\begin{corollary}[Multiple Winners Bidding with Zero Uncertainty]
\label{price-mu}

In equilibrium, in an unbounded pool with zero uncertainty, with $\mu < \lambda$ winners per round, a bidder with value $x$ bids consistently and repeatedly according to the bidding function $b(x)$
\begin{align*}
 b(x) &= \left\{  \begin{array}{ll}
x & x < X_{(\lambda / \mu)} \\
X_{(\lambda / \mu)} & x > X_{(\lambda / \mu)}
\end{array} \right.
\end{align*}
\end{corollary}

\begin{corollary}[Price in a Multiple Winners Auction]
\label{price}
In equilibrium, with zero uncertainty, with $\mu$ winners per round and a bidder arrival process with mean $\lambda > \mu$, the price paid for an item is $X_{(\lambda / \mu)}$.
\end{corollary}

The Markov chain for the single-winner model (with uncertainty. The zero-uncertainty version is not relevant to Bitcoin practice) derived in Section \ref{uncertainty} needs generalization to the multiple-winner model.

Let $G_\mu(z)$ be the PGF of the stationary distribution $\{p_n\} \equiv \{p_n^{\lambda [1 - F(x)], \delta, \mu}\}$, where the right-hand side highlights the fact that the distribution is of values above $x$, a Poisson subprocess with mean $\lambda [1 -  F(x)]$, and depends also on $\delta$ and $\mu$ (compare to the single-winner definition $\{p_n\} \equiv \{p_n^{\lambda [1 - F(x)], \delta}\}$ in Section \ref{winner}).

By definition, $G_\mu(z) := \sum\limits_{n=0}^\infty p_n z^n$. Generalizing \eqref{stationary-delta} to $\mu$ winners, $G_\mu(z)$ satisfies
\begin{align}
\label{stationary-mu}
G_\mu(z) = \Big(\sum\limits_{n=0}^{\mu-1} p_n + \sum\limits_{n=\mu}^\infty p_n [\delta + (1-\delta)z]^{n-1}\Big) \phi(z)
\end{align}

To generalize Observation \ref{W-ob}, there are at most $\mu$ winners every round. Given $p_n = p_n^{(\delta, \lambda[1-g])}$, the probability for $\mu$ winners with percentile $\leq g$ is $p_0$, the probability for $\mu-1$ such winners is $p_1$, etc.. If $W(g)$, the winners' distribution is the fraction of winners that have percentile at most $g$, we must have 
\begin{align}
\label{multi-W}
W(g) = \sum\limits_{j=0}^{\mu-1} (1 - j/\mu) p_j
\end{align}

\section{Price Announcements with Bidder Uncertainty}

In an unending auction with uncertainty, price announcements influence bidder strategies, adding a layer of complexity to equilibrium analysis that may render it intractable. In contrast, auctions without uncertainty have a simpler structure where price announcements do not affect strategies.

In the classical model of finite sequential auctions, price announcements do not impact strategies. This can be attributed to the fact that the winner distribution, conditional on price announcements, is proportional (in both c.d.f. and p.d.f.) to the unconditioned winner distribution. This constant proportion cancels out in the first-order conditions for equilibrium, a result that, ultimately, hinges on the mutual independence of bidder values.

As noted in footnote \ref{g}, our method for calculating equilibrium diverges from the standard approach, introducing a dependence of strategies on price announcements. However, the primary cause lies elsewhere. Specifically, it arises from the fact that Markov chains, and the winner values in our stationary distribution, exhibit autocorrelation. This implies that the winner values $w_n$ and $w_{n+k}$ from rounds $n$ and $n+k$, respectively, are not mutually independent.

A normalized autocovariance can be calculated between these values, which, like the Pearson correlation coefficient, is a pure number within $[-1, 1]$. This autocovariance decays to zero as $k$ increases, making the assumption of independence for distant rounds a reasonable approximation. However, for small $k$, the coefficients remain positive and significant.

\end{document}